\documentstyle[12pt]{article}
\begin{document}
\input amssym.def
\input amssym
\renewcommand{\refname} {Bibliography}
\centerline{\bf Approximate Solution of the Representability Problem }
\bigbreak
\centerline {A. I. Panin}
\bigbreak
\centerline{ Chemistry Department, St.-Petersburg State University,}
\centerline {University prospect 2, St.-Petersburg 198504, Russia }
\centerline { FAX: (07)--(812)--428--69--39;   e-mail: andrej@AP2707.spb.edu }

\bigbreak

{\bf ABSTRACT: }{\small Approximate solution of the ensemble
representability problem for density operators of arbitrary order is
obtained. This solution is closely related to the ``Q condition'' of A. J.
Coleman.  The representability conditions are formulated in orbital
representation and are easy to use. They are tested numerically on the
base of CI calculation of simple atomic and molecular systems. General
scheme of construction of the contraction operator right inverses is
proposed and the explicit expression for the right inverse associated with
the expansion operator is derived as an example. Two algorithms for direct
2-density matrix determination are described.}

{\bf Key words: }{\small representability problem, reduced density
operators, electron correlation.}

\bigbreak
\centerline{\bf 1.Introduction}
\bigbreak
In the last decade a number of Post HF methods of high precision have been
developed \cite {Schleyer}. It is unlikely, however, that in the nearest
future these methods in their present form have any chance to be applied
for electronic structure calculations of extensive molecular and
crystalline systems. Instead much more simple DFT methods \cite {Parr}
based on the work of Kohn \cite {Kohn} are becoming the main tool for
calculations of such systems.

There exists an approach in density functional theory which progress had
been restricted by serious mathematical problems connected with the
so-called representability property of 2-electron density. After the
famous theorem of Coleman \cite {Coleman-1} who solved the
representability problem for 1-density operators, there appeared a number
of papers with attempts to generalize Coleman's result to treat q-density
operators (for $q\ge 2$) \cite {Coleman-2, Kuhn, Kummer,
Erdahl, Mestechkin, Harriman}. These attempts, however, did not end in
results of practical importance. It even became accepted that the problem
is far too complicated to find any applications for electronic structure
calculations.

An alternative attempt to use reduced density matrices for molecular
calculations is developed in \cite {Mazziotti} (and references therein).

In present work the necessary conditions for Fermion representability
closely related to the Q condition of Coleman \cite {Coleman-3,Coleman-4}
are obtained for the most general case. Our technique is essentially
finite-dimensional and extensively use set-theoretical manipulations,
enumerative combinatorics, and multilinear algebra. For these reasons we
found it convenient to change notations that seems to be considered as
traditional in literature on the representability problem. Namely, capital
Roman letters are normally used to denote subsets of the spin-orbital
index set $N$ and/or the orbital index set M. Number of elements (indices)
in index sets are usually denoted by small Roman letters, e.g. $|N|=n$,
$|M|=m$, etc. With such notations it seems consistent to use small Roman
lettes for the number of electrons and for the current density matrix
order that are just the numbers of elements in the relevant spin-orbital
index sets. Throughout this paper $n$ is the number of spin orbitals, m is
the number of orbitals, p stands for the number of electrons, and q is the
density matrix order.

The following results described  in present paper we consider as novel:

Explicit expression for $A(n,p,q)$ operator is derived and its
connection with Coleman's $Q$ operator in the cases $q=1$ and $q=2$
is established;

It is proved that $A(n,p,q)$ in invertable and explicit analytic
expression for its inverse is obtained;

Distance function for the convex body giving an exterior approximation
of the set of representable density operators is obtained;

Analytic expression for the right inverse of the contraction operator
associated with the expansion operator is derived;

Description of the exterior approximation for the set of all representable
matrices is obtained in orbital basis for many electron systems with fixed
total spin projection;

Two algorithms for direct  determination of approximate 2-density
matrix are developed.

\newpage

\centerline{\bf 2.Contraction and Expansion Operators}
\bigbreak

Let ${\cal F}_{n,1}$ be one-electron Fock space spanned by an orthonormal
set $(\psi_i)$ of $n$ molecular {\sl spin-orbitals}. Electronic Fock space
is defined as

$${\cal F}_n=\bigoplus\limits_{p=0} {\cal F}_{n,p} \eqno(1)$$

where

$${\cal F}_{n,p}=\bigwedge\limits^p{\cal F}_{n,1}\eqno(2)$$
$${\cal F}_{n,0}={\Bbb C} \eqno(3)$$
where $\Bbb C$ is the field of complex numbers.

``Determinant'' basis vectors of the Fock space are conveniently labelled
by subsets of the spin-orbital index set $N$: for any $R\subset N$ the
corresponding basis determinant will be denoted by $|R\rangle$.

Creation-annihilation operators associated with spin-orbital index $i$ are
defined by the following relations

$$a_i^{\dag}|R\rangle=(1-\delta_{i,R})(-1)^{\epsilon}|R\cup i\rangle
\eqno(4a)$$
$$a_i|R\rangle=\delta_{i,R}(-1)^{\epsilon}|R\backslash i\rangle\eqno(4b)$$

where

$$\delta_{i,R}=\cases{1,&if $i\in R$\cr
                      0,&if $i\notin R $\cr}\eqno(5)$$

may be considered as a possible generalization of the Kronecker $\delta$
symbol, and

$$\epsilon=|\{ 1,2,\ldots,i-1\} \cap R|\eqno(6)$$

is the sign counter.

Let us introduce step-down and step-up (super)operators acting on the
operator space ${\cal F}_n\otimes {\cal F}_n^*$:

$$c_{ij}:z \to a_iza_j^{\dag}\eqno(7a)$$
$$u_{ij}:z \to a_i^{\dag}za_j\eqno(7b)$$

where $z$ is an arbitrary operator over ${\cal F}_n$. If $z\in {\cal
F}_{n,r}\otimes {\cal F}^*_{n,s} $ then $c_{ij}(z)\in {\cal
F}_{n,r-1}\otimes {\cal F}^*_{n,s-1}$ and $u_{ij}(z)\in {\cal
F}_{n,r+1}\otimes {\cal F}^*_{n,s+1}$.

The operator space ${\cal F}_n\otimes {\cal F}_n^*$ may be equipped with
the trace inner product

$$(z|t)= Tr(z^{\dag}t)\eqno(8)$$

It is easy to see that the step-down and step-up operators are mutually
conjugate with respect to this inner product.

Now we can define the contraction  and expansion  as

$$c=\sum\limits_{i}c_{ii}\eqno(9a)$$
$$u=\sum\limits_{i}u_{ii}\eqno(9b)$$

Their restriction to the $p$-electron section of the operator space is
identical (up to the nonessential combinatorial prefactor) to the commonly
used contraction and expansion.

Let $z_p\in {\cal F}_{n,p}\otimes {\cal F}_{n,p}^*$ be some p-electron
operator. It is expanded via basis ``determinant generators'' as

$$z_p=\sum\limits_{R,S}^{(p)}Z_{RS}|R\rangle \langle S|\eqno(10)$$

where the upper summation index $(p)$ indicates that the sum is taken over
all  $p$-element subsets of the spin-orbital index set $N$.  It can be
shown \cite {Panin} that

$$c^k(z_p)=k!\sum\limits_{R,S}^{(p)}Z_{RS}\sum\limits_{K\subset R\cap S}
^{(k)}(-1)^{|(R\Delta S)\cap \Delta_K|}|R\backslash K\rangle\langle S
\backslash K|\eqno(11)$$

The definition and properties of the set- theoretical operation $\Delta$
are collected in Appendix A.

Let us consider the electronic Hamiltonian of the form

$$H=\sum\limits_{i,j}\langle i|h|j\rangle a_i^{\dag}a_j+
{1\over 2}\sum\limits_{i,j,k,l}\langle ij|{1\over{r_{12}}}|kl\rangle
a_i^{\dag}a_j^{\dag}a_la_k\eqno(12)$$

which is obviously a particle number concerving operator: $H{\cal F}_{n,p}
\subset {\cal F}_{n,p}$ for any $p=0,1,2,\ldots$
The energy functional corresponding to some $p$-electron system is defined
as

$$E(z_p)=Tr(Hz_p)\eqno(13)$$

and is a linear mapping from ${\cal F}_{n,p}\otimes {\cal F}_{n,p}^*$ to
the field of complex numbers. In the particular case of a pure
$p$-electron state $z_p=|\Psi \rangle\langle \Psi|$ Eq.(13) reduces to the
standard average energy expression.  Using specific form of the electronic
Hamiltonian, it is possible to contract the energy domain and redefine
energy in terms of $q$-electron operators $(q\ge 2)$.  Introducing reduced
Hamiltonian

$$H_{p\to q}=\frac{q-1}{p-1}\sum\limits_{i,j}\langle i|h|j
\rangle a_i^{\dag}a_j+
{1\over 2}\sum\limits_{i,j,k,l}\langle ij|{1\over{r_{12}}}|kl\rangle
a_i^{\dag}a_j^{\dag}a_la_k\eqno(14)$$

we can rewrite the energy expression (13) in the form

$$E(z_p)=\frac{{p\choose 2}}{{q\choose 2}}
Tr\biggl(H_{p\to q}\frac{q!}{p!}c^{p-q}(z_p)\biggr)
\eqno(15)$$

The unknown operator on the right-hand side of this equation is q-electron
one with the unit trace. Therefore, if we were interested in the energy
functionals on $ {\cal F}_{n,p}\otimes {\cal F}_{n,p}^* $ we could easily
redefine them on ${\cal F}_{n,q}\otimes {\cal F}_{n,q}^*$ with $2\le q\le
p$. Unfortunately, the actual situation is much more complicated. Indeed,
physically relevant are the so-called {\sl density operators} that are
distinguished from general linear operators by the following three
properties:

(i) Positive semidefinitness;

(ii) Hermiteancy;

(iii) Unit trace.

The set of all $p$-electron density operators will be denoted by ${\cal
E}_{n,p}$.  The set ${\cal E}_{n,p}$ may be characterized both
analytically and parametrically.  Indeed, axioms (i)-(iii) give us
analytic description of ${\cal E}_{n,p}$ in terms of linear equalities and
inequalities. On the other hand, it is not difficult to recognize that
${\cal E}_{n,p}$ is a convex hull of the so-called pure states:

$$t_p\in {\cal E}_{n,p}\leftrightarrow t_p=\sum\limits_{i=1}^{n\choose p}
{\lambda}_i
|{\Psi}_i\rangle\langle{\Psi}_i|\eqno(16)$$

where

$${\lambda}_i\ge 0, \qquad i=1,2,\ldots,{n\choose p}\eqno(17)$$
$$\sum\limits_{i=1}^{n\choose p}{\lambda}_i=1\eqno(18)$$

and ${\Psi}_i\in {\cal F}_{n,p}$.

It can be easily verified also that

$$\frac{q!}{p!}c^{p-q}{\cal E}_{n,p}\subset {\cal E}_{n,q}\eqno(19)$$

The set

$${\cal W}_{n,p,q}=\frac{q!}{p!}c^{p-q}({\cal E}_{n,p})\eqno(20)$$

is a proper subset of ${\cal E}_{n,q}$ and is called {\sl the set of all
$p$--representable density operators of order $q$}. Unfortunately, in
general case the structure of these sets seems to be so complicated that
numerous attempts to find their constructive description have not lead to
practically useful results.

Taking some inner point of ${\cal W}_{n,p,q}$, we can try to construct
functions that determine the distance from a fixed inner point of
${\cal W}_{n,p,q}$ to its border ${\partial}{\cal W}_{n,p,q}$ along some
chosen direction. There exists natural {\sl central inner point} of the
convex sets ${\cal E}_{n,q}$ and ${\cal W}_{n,p,q}$ (the normalized
identity operator):

$$ w_q=\frac{1}{{n\choose q}}\sum\limits_{K\subset
N}^{(q)}|K\rangle\langle K|\eqno(21)$$

where $N$ is the set of spin-orbital indices, $|N|=n$. With respect to
this central point the energy functional may be rewritten as

$$E(t_q,\varepsilon(t_q))=\frac{{p\choose 2}}{{q\choose 2}}
Tr(H_{p\to q}\lbrace w_q+
\varepsilon(t_q)\lbrack t_q-w_q\rbrack \rbrace) \eqno(22)$$

where $\varepsilon(t_q)$ determines the distance from $w_q$ to the border
of ${\cal W}_{n,p,q}$ along the direction $t_q-w_q$.  A technique of
distance function $\varepsilon(t_q)$ values calculation involving
minimization of a certain auxiliary function can be developed. In our
opinion, however, such an approach may be even more complicated than the
commonly used CI one.  Anyway, in present paper we will not try to follow
this route.

\newpage
\centerline{\bf 3. Exterior  Approximation  for the Convex Set }
\centerline {\bf of Representable Density Operators.}
\bigbreak
First of all let us analyze one of two cases where the ensemble
representability problem admits constructive solution \cite {Erdahl}. To
this end it is convenient to introduce a new basis in the operator space
that differs from the basis of the determinant generators only by the
phase transformation \cite {Panin}:

$$e_L^{IJ}=(-1)^{|(I\cup J)\cap \Delta_L|}|I\cup L\rangle \langle J\cup L|
\eqno(23)$$

where $I\cap J=\emptyset $.  From this definition and Eq.(11) it readily
follows that

$$c^{p-q}e_L^{IJ}=(p-q)!\sum\limits_{K\subset L}^{(q)}e_K^{IJ}\eqno(24)$$
$$u^{p-q}e_K^{IJ}=(p-q)!\sum\limits_{L\supset K}^{(q)}e_L^{IJ}\eqno(25)$$

Over the finite-dimensional electronic Fock space there exists an
involution defined as

$$\Bbb I|R\rangle=|N\backslash R\rangle\eqno(26)$$

Its restriction to the p-electron section of the Fock space is an
isomorphic mapping from ${\cal F}_{n,p}$ to ${\cal F}_{n,n-p}$ and

$${\Bbb I}_p {\Bbb I}_{n-p}=id_{{\cal F}_{n,n-p}}\eqno(27a)$$
$${\Bbb I}_{n-p} {\Bbb I}_p=id_{{\cal F}_{n,p}}\eqno(27b)$$
where $id_{{\cal F}_{n,n-p}}$ and $id_{{\cal F}_{n,p}}$ are the indentity
operators over ${{\cal F}_{n,n-p}}$ and ${{\cal F}_{n,p}}$,
correspondingly.  With each vector $|\Psi\rangle \in {\cal F}_{n,p}$ a
semilinear functional $\langle \Psi|:{\cal F}_{n,p} \to \Bbb C$ is
associated and

$$\langle \Psi|{\Bbb I}_{n-p}=\langle {\Bbb I}_p\Psi|\eqno(28)$$

where $\langle \Psi|{\Bbb I}_{n-p}$ is a composition of mappings ${\cal
F}_{n,n-p}\stackrel{{\Bbb I}_{n-p}}\longrightarrow {\cal F}_{n,p}
\stackrel{\langle \Psi|}\longrightarrow \Bbb C$.

For each pair $(n-p,p)$ let us put

$${\Bbb I}_{n-p,p}:z_p \to {\Bbb I}_p z_p {\Bbb I}_{n-p}\eqno(29)$$

where $z_p$ is an arbitrary $p$-electron operator. It is immediately clear
that ${\Bbb I}_{n-p,p}$ is an isomorphic mapping of ${\cal F}_{n,p}\otimes
{\cal F}_{n.p}^*$ onto ${\cal F}_{n,n-p}\otimes {\cal F}_{n,n-p}^*$.
Simple manipulations lead to the conclusion that

$${\Bbb I}_{n-p,p}({\cal E}_{n,p}) = {\cal E}_{n,n-p}\eqno(30)$$

The effect of ${\Bbb I}_{n-p,p}$ on the basis operators (23) is given by

$${\Bbb I}_{n-p,p}(e_L^{IJ})=(-1)^{\alpha_{IJ}}e_K^{JI}\eqno(31)$$

where $\alpha_{IJ}=|(I\cup J)\cap \Delta_{N\backslash (I\cup J)}|,{\
}\mbox {and} {\ } K=N\backslash (I\cup J\cup L)$. Note that the sign
prefactor on the right-hand side of Eq.(31) depends on the index set
$I\cup J$ but not on the number of elements in it.

Let us introduce certain symmetric combinations of $q$-electron and
$p$-electron basis operators:

$$w^{IJ}_{p\downarrow q}(L)=\frac{1}{{p-s\choose q-s}}\sum\limits
_{K\subset L}^{(q-s)}e^{IJ}_K\eqno(32)$$
$$w^{IJ}_{q\uparrow p}(K)=\frac{1}{{n-q-s\choose n-p-s}}\sum\limits
_{L\supset K}^{(p-s)}e^{IJ}_L\eqno(33)$$

where $I\cap J=\emptyset,{\ }|I|=|J|=s,{\ }K,L\subset N\backslash (I\cup
J), {\ }|L|=p-s,{\ }$ and $|K|=q-s$.

Using definitions (32)and (33), we can rewrite Eqs.(24) and (25) in the
form

$$\frac{q!}{p!}c^{p-q}e^{IJ}_L={{p-s\choose q-s}\over {p\choose q}}
w^{IJ}_{p\downarrow q}(L)\eqno(34)$$
$$\frac{(n-p)!}{(n-q)!}u^{p-q}e^{IJ}_K=\frac{{n-q-s\choose n-p-s}}
{{n-q\choose p-q}}w^{IJ}_{q\uparrow p}(K)\eqno(35)$$

Let us suppose that $p+q=n$. In this case the contraction operator is an
isomorphic mapping \cite {Erdahl, Panin} which means that the set of
vectors (32) is a basis set of the q-electron operator space. We can
therefore define nondegenerate linear operator in the following manner:

$$A(n,p,q)w^{IJ}_{p\downarrow q}(L)=(-1)^{\alpha_{IJ}}{{p\choose q}
\over {p-s\choose q-s}}e^{JI}_{N\backslash (I\cup J\cup L)}\eqno(36)$$

Comparing Eqs.(31),(35),and(36) immediately leads to the conclusion that
for $p+q=n$

$$A(n,p,q)\frac{q!}{p!}c^{p-q}={\Bbb I}_{n-p,p}\eqno(37)$$

and, consequently,

$${A(n,p,q){\cal W}_{n,p,q}={\cal E}_{n,q}}\atop {(p+q=n)}\eqno(38)$$

It seems pertinent to note that operator satisfying Eq.(38) is not unique.
We can as well take any composition of $A(n,p,q)$ with operator leaving
the set ${\cal E}_{n,q}$ invariant. For example, the transposition

$$\tau: |R\rangle\langle S|\to |S\rangle\langle R|\eqno(39)$$

possesses this property and we can use $\tau A(n,p,q){\ }$ instead of
$A(n,p,q)$. In contrast to the operator $A(n,p,q)$ the operator $\tau
A(n,p,q)$ is conveniently block-diagonal: $$\tau
A(n,p,q)=\bigoplus_{I,J}\tau A^{IJ}(n,p,q)\eqno(40)$$ where
$A^{IJ}(n,p,q)$ is the restriction of $A(n,p,q)$ on the subspace of
$p$-electron operator space spanned by the basis operators (23) with fixed
$I,J$.

After complicated combinatorial manipulations (closely related to the
famous inclusion-exclusion principle) the explicit matrix representation
of the operator $\tau A(n,p,q)$ can be obtained (see Appendix B)

$$\tau A(n,p,q)e^{IJ}_K=
(-1)^{\alpha_{IJ}} \frac{{p\choose q}}{{p-s\choose q-s}{n-p\choose q}}
\sum\limits_{K'\subset N\backslash (I\cup J)}^{(q-s)}(-1)^{|K\cap K'|}
\frac{{p+|K\cap K'|-q-1\choose {|K\cap K'|}}}{{q-s\choose |K\cap K'|}}
e^{IJ}_{K'}\eqno(41)$$
By an abuse of notation, we will use the symbol $A(n,p,q)$ for the
operator defined by Eq.(41).

Thus, in the case $p+q=n$ the convex set ${\cal W}_{n,p,q}$ can be
explicitly characterized in several equivalent ways:

as  the pre-image of ${\cal E}_{n,q}$ with respect to $A(n,p,q)$
$$ {\cal W}_{n,p,q}=A^{-1}(n,p,q){\cal E}_{n,q};\eqno(42a)$$

as the set of solutions of the system
$$\cases{ \langle{\Phi}_q|A(n,p,q)t_q|{\Phi}_q\rangle\ge 0, {\Phi_q}\in {\cal F}_{n,q} \cr
           Tr(t_q)=1;\cr}\eqno(42b)$$

as a  convex  body with the distance function
$$\varepsilon(t_q)=\frac{1} { 1-
{{n\choose q} {\min\limits_{||{\Phi}_q||=1}\langle{\Phi}_q|A(n,p,q)
t_q|{\Phi}_q\rangle} }} \eqno(42c)$$
where ${\Phi}_q\in {\cal F}_{n,q}$ and $Tr(t_q)=1$.

Let us try to generalize the approach described to handle the case
$p+q<n$.  The first idea coming to mind is to use the operator defined by
Eq.(41) in general case. This operator is nondegenerate for $n\ge p \ge q,
{\ }p+q\le n$ and explicit analytic expression for its inverse can be
obtained (see Appendix C). It can also be shown that

$$A(n,p,q){\cal W}_{n,p,q}\subset {\cal E}_{n,q}\eqno(43)$$

(see Appendix D).  From this inclusion and positive semidefiniteness of
the reduced density operators it readily follows that the convex set
$${\cal V}_{n,p,q}=A^{-1}(n,p,q){\cal E}_{n,q}\cap {\cal
E}_{n,q}\eqno(44)$$ is an exterior approximation for ${\cal W}_{n,p,q}$.
Simple combinatorial manipulations lead to the conclusion that
$$A(n,p,q)w_q=w_q$$ that is ${\cal V}_{n,p,q},{\cal W}_{n,p,q},{\ }$ and
${\ }{\cal E}_{n,q}$ share the same central point.

Thus, we have constructed the compact convex set that may serve as a
certain exterior approximation for the set ${\cal W}_{n,p,q}$ and coincide
with ${\cal W}_{n,p,q}$ in the particular case $p+q=n$. It is not
difficult to demonstrate, however, that in the case $q=1$ this set differs
from ${\cal W}_{n,p,1}$.  The reason is the unitary noninvariance of
$A(n,p,q)$.

It is well-known that the contraction operator is invariant with respect
to the induced unitary transformations of the type

$${\rho }_k(u)=(\bigwedge\limits ^k u)\otimes (\bigwedge\limits ^k
u^{\dag}), {\ } u\in U({\cal F}_{n,1})\eqno(45)$$
that is
$$ c{\rho}_k(u)t_k={\rho}_{k-1}(u)ct_k\eqno(46)$$

for any $t_k\in {\cal F}_{n,k}\otimes {\cal F}_{n,k}^* $.  Taking into
account obvious unitary invariance of the set ${\cal E}_{n,k}$, we can
conclude that

$${\rho}_q(u){\cal W}_{n,p,q}={\cal W}_{n,p,q}$$
for any $u\in U({\cal F}_{n,1}) $.

Direct calculations show that to restore the desired unitary invariance of
${\cal V}_{n,p,1}$ it is sufficient to modify the phase prefactors in
definition (41)(see Appendix E).

Let us introduce the phase transformation

$$\nu: |R\rangle \langle S| \to (-1)^{|R\Delta S)\cap \Delta_N|}|R\rangle
\langle S|$$
It leaves invariant the sets ${\cal E}_{n,q}$ and, consequently, we can
replace the operator $A(n,p,q)$ by $\nu A(n,p,q)$. Matrix representation
of the last operator differs from Eq.(41) only by the sign prefactor that
should be taken equal to $(-1)^{|(I\cup J)\cap {\Delta}_{(I\cup J)}|}$.
Hereafter, only the operator $\nu A(n,p,q)$ will be under consideration
and, by an abuse of notation, we will use for it the same symbol
$A(n,p,q)$.

In the particular case $q=1$ we have

$$A(n,p,1)e^{\emptyset \emptyset}_k=\sum\limits_{l=1}^n
{{1-p\delta_{kl}}\over {n-p}}e^{\emptyset \emptyset}_l\eqno(47a)$$
$$A(n,p,1)e^{ij}_{\emptyset}=-{p\over n-p}e^{ij}_{\emptyset}\eqno(47b)$$

Using the unitary invariance of $A(n,p,1)$ we can state that $t_1\in {\cal
V}_{n,p,1}$ if and only if $u^{\dag}t_1u\in {\cal V}_{n,p,1}$ where $u$ is
the unitary transformation diagonalizing density operator $t_1$.But for
diagonal $t_1$, with the aid if Eq.(47a), we can obtain the equivalence

$$ t_1=\sum\limits_{i=1}^n t^{\emptyset \emptyset}_ie^{\emptyset
\emptyset}_i\in {\cal V}_{n,p,1} \Leftrightarrow
\cases {0\le t_i^{\emptyset \emptyset}\le \frac{1}{p}\cr
Tr(t_1) =1}\eqno(48)$$

that immediately implies ${\cal V}_{n,p,1}={\cal W}_{n,p,1}$(consequence
of the diagonal version of the famous Coleman's theorem \cite
{Coleman-1}).

For the cases q=1 and q=2 it is easy to recast operator $A(n,p,q)$ in more
invariant form. We have

$${n-p\choose 1}A(n,p,1)t_1=-{p\choose 1}t_1+Tr(t_1)id_{{\cal
F}_{n,1}}\eqno(49)$$

and

$${n-p\choose 2}A(n,p,2)t_2={p\choose 2}t_2-p(t_1\wedge id_{{\cal
F}_{n,1}} + id_{{\cal F}_{n,1}}\wedge t_1) +Tr(t_2)\bigwedge\limits^2
id_{{\cal F}_{n,1}}\eqno(50)$$

where $ id_{{\cal F}_{n,1}}$ is the identity operator over ${\cal
F}_{n,1}$and $t_1={1\over 2}ct_2$ is $2 \to 1$ contracion of $t_2$. The
operator (50) is similar but not identical to the Coleman's operator Q
\cite {Coleman-3, Coleman-4}. Operator  $A(n,p,2)$ differs from $Q$ in
a rather delicate manner. Namely, ${n-p\choose 2}A(n,p,2)t_2$ includes
symmetric term $p(t_1\wedge id_{{\cal F}_{n,1}}
+ id_{{\cal F}_{n,1}}\wedge t_1)$ whereas $Qt_2$ contains the term
$2p(t_1\wedge id_{{\cal F}_{n,1}})$ which, in our opinion, is not
correctly defined. The situation here is completely analogous to that in the
angular momentum theory. For example, for two-electron systems one-electron
operator of momentum projecton $j_z$ has no sense in itself.
Only the total operator $J_z=j_z\wedge id_{{\cal F}_{n,1}}+
id_{{\cal F}_{n,1}}\wedge j_z$ is of mathematical and physical meaning
and it can not be replaced by $2(j_z\wedge id_{{\cal F}_{n,1}})$.

The sets ${\cal V}_{n,p,q}$ may be explicitely described in several
equivalent ways:

As the intersection of ${\cal E}_{n,q}$ with its pre-image with respect to
$A(n,p,q)$(see Eq.(44));

As the set of solutions of the system
$$\cases{ \langle{\Phi}_q|t_q|{\Phi}_q\rangle\ge 0,\cr
           \langle{\Phi}_q|A(n,p,q)t_q|{\Phi}_q\rangle\ge 0\cr
           Tr(t_q)=1\cr}\eqno(51a)$$
where ${\Phi}_q\in {\cal F}_{n,q};$

As a  convex  body with the distance function

$$\varepsilon(t_q)=min\lbrace\frac{1}{1-{{n\choose
q}{\min\limits_{||{\Phi}_q||=1}
\langle{\Phi}_q|t_q|{\Phi}_q\rangle}}},
\frac{1}{ 1-
{{n\choose q} {\min\limits_{||{\Phi}_q||=1}\langle{\Phi}_q
|A(n,p,q)t_q|{\Phi}_q\rangle} }}
\rbrace\eqno(51b)$$

where ${\Phi}_q\in {\cal F}_{n,q}$ and $Tr(t_q)=1$.

From the well-known properties of the distance functions (see, e.g., \cite
{Leicht}) it follows that $t_q\in \partial {\cal V}_{n,p,q}$ if and only
if $\varepsilon(t_q)=1$.  But from Eq.(51b) it is easy to see that
$\varepsilon(t_q)=1$ if and only if either $t_q$ or $A(n,p,q)t_q$ has at
least one zero eigenvalue.  In the particular case of pure $p$-electron
determinant state $|R\rangle \langle R|$ the corresponding $q$-density
matrix

$$t^{HF}_q(R)=\frac{1}{{p\choose q}}\sum\limits_{K\subset R}^{(q)}
|K\rangle\langle K|\eqno(52)$$

has ${n\choose q}-{p\choose q}$ zero eigenvalues and, being obviously
representable, belongs to $\partial {{\cal V}_{n,p,q}}\cap \partial {{\cal
W}_{n,p,q}}$.

Direct but somewhat tedious combinatorial calculations  lead to the
important commutation relation  of the operator $A(n,p,q)$ with the
contraction operator (see Appendix F):

$$A(n,p,q-1)c=cA(n,p,q)\eqno(53)$$

Now we can collect the most important properties of  the convex
set ${\cal V}_{n,p,q}$:

(1) {\sl It gives an exterior approximation  for the set} ${\cal W}_{n,p,q}$
{\sl of representable density operators};

(2) {\sl From Eq.(53) it follows that}
$${\ } {\cal V}_{n,p,q}={\cal W}_{n,p,q} \ for\  q=n-p\
and\  q=1;\eqno(54a)$$
$${\ }\frac{1}{q}c{\cal V}_{n,p,q}\subset {\cal V}_{n,p,q-1} \ for\  any\
q\le p\eqno(54b)$$

(3) {\sl The set ${\cal V}_{n,p,q}$ is invariant with respect to
transformations induced by unitary 1-electron ones (see Eq.(45))};

(4) {\sl All Hartree-Fock densities belong to the border of} ${\cal
V}_{n,p,q}$;

(5) {\sl Any representable $q$-density matrix $t_q$ such that either $t_q$
or its image with respect to $A(n,p,q)$ possess zero eigenvalue belongs to
the border} of ${\cal V}_{n,p,q}$.

Density matrix of order $q$ associated with pure $p$-electron state  practically
always possesses property (5).  Indeed, if there exits one-electron basis
with at least one molecular spin-orbital having occupancy 0 (virtual) or 1
(inactive) then block-diagonal matrix $t_q \oplus A(n,p,q)t_q$ necessarily
has zero or very small eigenvalue. Practice of {\sl ab initio} calculations shows
that even in bases of moderate size among natural spin-orbitals there present
inactive and/or virtual ones. We may therefore hope that the direct energy
optimization on the set ${\cal V}_{n,p,q}$ will lead to reasonable results.

For readers who prefer numerical arguments instead of abstract
mathematical ones, in Section 5 the results of relevant calculations
testing property (5) of the set ${\cal V}_{n,p,2}$ are presented.

\newpage
\centerline{\bf 4. Sections of the Contraction Operator}
\bigbreak
The contraction operator $\frac{q!}{p!}c^{p-q}$ is linear {\sl surjective}
mapping from ${\cal F}_{n,p}\otimes {\cal F}_{n,p}^*$ on ${\cal
F}_{n,q}\otimes {\cal F}_{n,q}^*$.  By definition, global section of the
contraction operator (or its right inverse) is linear {\sl injective}
mapping $\pi_{q\uparrow p}$ from ${\cal F}_{n,q}\otimes {\cal F}_{n,q}^*$
to ${\cal F}_{n,p}\otimes {\cal F}_{n,p}^*$ such that

$$\frac{q!}{p!}c^{p-q}\pi_{q\uparrow p}=
id_{{\cal F}_{n,q}\otimes {\cal F}_{n,q}^*}\eqno(55)$$

Here we describe a certain general scheme  of construction of
the contraction operator sections and illustrate this scheme on
example of section associated with the expansion operator..

First of all we should find an injective linear mapping
${\gamma}_{q\uparrow p}$ of the form

$${\gamma}_{q\uparrow p}(e_K^{IJ})=\sum\limits_
{L\in {\mathcal L}_K^{IJ}}^{(p-s)}e^{IJ}_Lf^{IJ}(L,K)\eqno(56)$$

where ${\mathcal L}_K^{IJ}$ is a certain set of $(p-s)$-element subsets
from $N\backslash (I\cup J)$. Injectiveness of the mapping (56) means that
vectors ${\gamma }_{q\uparrow p}(e^{IJ}_K)$ are lineary independent.  Then
we should calculate

$$\frac{q!}{p!}c^{p-q}{\gamma}_{q\uparrow p}(e^{IJ}_K)=
\sum\limits_{K'\subset N\backslash (I\cup J)}^{(q-s)}e^{IJ}_{K'}
G^{IJ}_{K'K}\eqno(57)$$

where we  introduced (in general non-symmetric) matrix $G(n,p,q)$ with
matrix elements

$$G^{IJ}_{K'K}=\frac{1}{{p\choose q}}
\sum\limits_{{K'\subset L \in {\mathcal L}^{IJ}_K}}^{(p-s)}
f^{IJ}(L,K)\eqno(58)$$

If this matrix is invertable then the desired section of the contraction
operator may be written as

$${\pi}_{q\uparrow p}={\gamma}_{q\uparrow p}G^{-1}(n,p,q)\eqno(59)$$

As an example, let us consider the section associated with the expansion
operator (see Eq.(35)): ${\gamma}_{q\uparrow
p}=\frac{(n-p)!}{(n-q)!}u^{p-q}$.  In this case in Eq.(56) we have
${\mathcal L}^{IJ}_K=\{L\subset N\backslash (I\cup J): L\supset K\}$, and
$f^{IJ}(L,K)=\frac{1}{{n-q\choose p-q}}$. First of all we should ascertain
that the expansion operator is injective. To this end it is sufficient to
show that the vectors (33) are lineary independent. Then we should
construct matrix G(n,p,q) and try to invert it. All relevant technical
details are collected in Appendix G. Here we just write down the final
explicit expression for the section under discussion:

$${\pi}_{q\uparrow p}(e^{IJ}_K)=
(-1)^{q-s}\frac{{p\choose q}}{{n-q-s\choose p-q}}
\sum\limits_{L\subset N\backslash (I\cup J)}^{(p-s)}e^{IJ}_L
(-1)^{|K\cap L|}\frac{{p-s-|K\cap L|-1\choose q-s-|K\cap L|}}
{{n-p-s\choose q-s-|K\cap L|}}\eqno(60)$$

For the case $p+q=n$ there exists the unique global section of the
contraction operator equal to its inverse (see Eqs.(29) and (41)):
${\pi}_{q\uparrow p}={\Bbb I}_{p,n-p}A(n,p,q)$. We can try to generalize
this last section to treat the case $p+q>n$. One of the ways is to define
vectors

$$v^{IJ}_K=(-1)^{q-s}\frac{{p\choose q}}{{p-s\choose q-s}{n-p\choose q}}
\sum\limits_{L\subset N\backslash (I\cup J)}^{(p-s)}e^{IJ}_L
(-1)^{|L\cap K|}\frac{{p-s-|K\cap L|-1\choose q-s-|K\cap L|}}
{{q-s\choose |K\cap L|}}\eqno(61)$$

and consider the mapping ${\gamma}_{q\uparrow p}(e^{IJ}_K)=v^{IJ}_K$.

Besides global sections satisfying Eq.(55) of interest are also sections
closely related to the concrete density matrix under consideration and
defined on a certain subspace of ${\cal F}_{n,q}\otimes {\cal F}_{n,q}^*$.
They can be used to restore the part of $p$-density matrix that survives
under contraction to the level q. Any progress in this direction would be
of great importance.

\newpage
\centerline{\bf 5. Density Operators in Orbital Representation}
\bigbreak

Let $(\varphi_i)$ be an orthonormal set of $m$ {\sl orbitals}. Following
Handy \cite {Handy} we identify $p$-electron determinants generated by
these MOs, with pairs of index sets (strings):

$$|R_{\alpha},R_{\beta}\rangle=a^{\dag}_{i_1\alpha}\ldots
a^{\dag}_{i_{p_{\alpha}}\alpha}a^{\dag}_{j_1\beta}\ldots
a^{\dag}_{j_{p_{\beta}}\beta}|\emptyset\rangle\eqno(62)$$
where $R_{\alpha}=1<i_1<\ldots<i_{p_{\alpha}}<m$,${\ }R_{\beta}=
1<j_1<\ldots <j_{p_{\beta}}<m$,  $p_{\alpha}+p_{\beta}=p$, and
$|\emptyset \rangle $ is the vacuum vector.

Split basis operators analogous to ones defined by Eq.(23) are

$$e_{(L_{\alpha},L_{\beta})}^{(I_{\alpha},J_{\alpha})(I_{\beta},J_{\beta})}=
(-1)^{\epsilon_1}|I_{\alpha}\cup L_{\alpha},I_{\beta}\cup L_{\beta}\rangle
\langle J_{\alpha}\cup L_{\alpha},J_{\beta}\cup L_{\beta}|\eqno(63)$$

where $I_{\alpha}\cap J_{\alpha}=I_{\beta}\cap J_{\beta}=\emptyset,$
$$\epsilon_1=|I_{\alpha}\cup J_{\alpha}|\times |L_{\beta}|_2+
|(I_{\alpha}\cup J_{\alpha})\cap \Delta_{L_{\alpha}}|+
|(I_{\beta}\cup J_{\beta})\cap \Delta_{L_{\beta}}|,$$

$$|L_{\beta}|_2=\cases{0,&if $|L_{\beta}|\equiv 0{\ } (mod{\ } 2)$\cr
                       1,&if $|L_{\beta}|\equiv 1{\ } (mod{\ } 2)$\cr}$$
and
$$\frac{q!}{p!}c^{p-q}
e_{(L_{\alpha},L_{\beta})}^{(I_{\alpha},J_{\alpha})(I_{\beta},J_{\beta})}=
\frac{1}{{p\choose q}}
\sum\limits_{k_{\alpha},k_{\beta} \atop {(k_{\alpha}+k_{\beta}=q-s})}
\sum\limits_{K_{\alpha}\subset L_{\alpha}}^{(k_{\alpha})}
\sum\limits_{K_{\beta}\subset L_{\beta}}^{(k_{\beta})}
e_{(K_{\alpha},K_{\beta})}^{(I_{\alpha},J_{\alpha})(I_{\beta},J_{\beta})}=$$

$$=\frac{{p-s\choose q-s}}{{p\choose q}}
w_{p\downarrow q}^{(I_{\alpha},J_{\alpha})(I_{\beta},J_{\beta})}
(L_{\alpha},L_{\beta})\eqno(64)$$

The operator $A(2m,p,q)$ acts on the basis vectors (63) as

$$A(2m,p,q)e_{(K_{\alpha},K_{\beta})}^{(I_{\alpha},J_{\alpha})
(I_{\beta},J_{\beta})}=
(-1)^{\epsilon_2} \frac{{p\choose q}}{{p-s\choose q-s}{2m-p\choose q}}
\sum\limits_{k_{\alpha},k_{\beta} \atop {(k_{\alpha}+k_{\beta}=q-s})}
\sum\limits_{K'_{\alpha}\subset {M\backslash (I_{\alpha}
\cup J_{\alpha})}}^{(k_{\alpha})}$$
$$\sum\limits_{K'_{\beta}\subset {M\backslash (I_{\beta}\cup
J_{\beta})}}^{(k_{\beta})}
(-1)^{|K_{\alpha}\cap K'_{\alpha}|+|K_{\beta}\cap K'_{\beta}|}
\frac{{p+|K_{\alpha}\cap K'_{\alpha}|+|K_{\beta}\cap K'_{\beta}|-q-1\choose
|K_{\alpha}\cap K'_{\alpha}|+|K_{\beta}\cap K'_{\beta}|}}{{q-s\choose
|K_{\alpha}\cap K'_{\alpha}|+|K_{\beta}\cap K'_{\beta}|}}
e_{(K'_{\alpha},K'_{\beta})}^{(I_{\alpha},J_{\alpha})
(I_{\beta},J_{\beta})}\eqno(65)$$

where $M$ is the orbital index set,
$s=|I_{\alpha}|+|I_{\beta}|=|J_{\alpha}|+|J_{\beta}|$, and

$$\epsilon_2=|I_{\alpha}\cup J_{\alpha}|\times |I_{\beta}\cup J_{\beta}|_2+
|(I_{\alpha}\cup J_{\alpha})\cap \Delta_{(I_{\alpha}\cup J_{\alpha})}|+
|(I_{\beta}\cup J_{\beta})\cap \Delta_{(I_{\beta}\cup J_{\beta})}|$$

Note that the sign counters $\epsilon_1$ and $\epsilon_2$ in these
formulas are determined by the initial {\sl spin orbital} index set
ordering and its current form corresponds to the split determinant
representation (62), where $\alpha $ indices always go first.

Let us consider p-electron states with a given value of the total spin
projection $M_S$.  The set of all such states is generated by the
determinants (62) with fixed
$|R_{\alpha}|=p_{\alpha},|R_{\beta}|=p_{\beta},$ since, by assumption,
$p_{\alpha}+p_{\beta}=p,$ and $p_{\alpha}-p_{\beta}=2M_S$. Contraction of
an arbitrary determinant generator $p-2$ times may lead to nonzero result
only if $|R_{\alpha}\cap R_{\beta}|+ |S_{\alpha}\cap S_{\beta}|\ge p-2$
which means that on the 2-electron level there may appear {\sl only three
types of generators}:

$$|r_1r_2,\emptyset \rangle \langle s_1s_2,\emptyset |$$
$$|\emptyset,r_1r_2 \rangle \langle \emptyset ,s_1s_2|$$
$$|r_1,r_2 \rangle \langle s_1,s_2|$$

As a result, of interest are block-diagonal 2-electron density operators
of the form
$$t_2=t_2^{\alpha}+t_2^{\beta}+t_2^{\alpha \beta}\eqno(66)$$
where
$$ t_2^{\alpha}=\sum\limits_{r_1<r_2\atop {s_1<s_2}}t^{\alpha}_{r_1r_2;s_1s_2}
|r_1r_2,\emptyset\rangle \langle s_1s_2,\emptyset|\eqno(66a)$$
$$ t_2^{\beta}=\sum\limits_{r_1<r_2\atop {s_1<s_2}}t^{\beta}_{r_1r_2;s_1s_2}
|\emptyset,r_1r_2\rangle \langle \emptyset,s_1s_2|\eqno(66b)$$
$$ t_2^{\alpha \beta}=\sum\limits_{r_1,r_2\atop {s_1,s_2}}
t^{\alpha \beta}_{r_1,r_2;s_1,s_2}
|r_1,r_2\rangle \langle s_1,s_2|\eqno(66c)$$

Calculation of traces of 2-density matrix components corresponding to some
$p$-electron wavefunction of fixed total spin projection gives

$${Tr(t_2^{\sigma})=\frac{{p_{\sigma}\choose 2}}{{p\choose 2}}}
\atop {(\sigma=\alpha,\beta)}\eqno(67a)$$

and

$$Tr(t_2^{\alpha \beta})=\frac{p_{\alpha}p_{\beta}}{{p\choose 2}}
\eqno(67b)$$

Eqs.(67a)-(67b) may be considered as additional restrictions  on
density matrix components in the course of energy optimization.

Using either general definition (65) or invariant operator (50) one can
show that the action of the operator $A(n,p,2)$ on arbitrary 2-density
matrix $t_2$ of block-diagonal form (66) results in matrix of the same
block-diagonal structure

$$d_2=A(n,p,2)t_2=d_2^{\alpha}+d_2^{\beta}+d_2^{\alpha \beta}\eqno(68)$$

and trace restrictions (67a)-(67b) imply the following restrictions on
the traces of spin-components of operator $d_2$:

$${Tr(d_2^{\sigma})=\frac{{m-p_{\sigma}\choose 2}}{{2m-p\choose 2}}}
\atop {(\sigma=\alpha,\beta)}\eqno(69a)$$
and
$$Tr(d_2^{\alpha \beta})=\frac{(m-p_{\alpha})(m-p_{\beta})}
{{2m-p\choose 2}} \eqno(69b)$$

Reduced Hamiltonian in the orbital representation is

$$H_{p\to 2}=\frac{1}{p-1}\sum\limits_{i,j=1}^mh_{ij}E_{ij}+
\frac{1}{2}\sum\limits_{i,j,k,l=1}^m(ij|kl)
\lbrack E_{ij}E_{kl}-\delta_{jk}E_{il}\rbrack\eqno(70)$$

where

$$h_{ij}=\int\limits_{\Bbb R^3}\varphi _i ^*(r_1)h\varphi
_j(r_1)dr_1\eqno(71)$$
$$(ij|kl)=\int\limits_{\Bbb R^3\times \Bbb R^3}\varphi _i^*(r_1)
\varphi _j(r_1)\frac{1}{r_{12}}\varphi _k^*(r_2)\varphi
_l(r_2)dr_1dr_2\eqno(72)$$

and

$$E_{ij}=E^{\alpha }_{ij}+E^{\beta }_{ij}=a^{\dag }_{i\alpha }a_{j\alpha }+
a^{\dag }_{i\beta }a_{j\beta }\eqno(73)$$

are the unitary group generators.

Standard but somewhat tedious algebraic manipulations lead to the following
general energy expression

$$E(t_2)={p\choose 2}\biggl[ \frac{1}{p-1}\sum\limits_{i,j=1}^m
\bigl[ \sum\limits_{k=1}^m(\bar t^{\alpha}_{ik;jk}+\bar t^{\beta}_{ik;jk}+
t^{\alpha \beta}_{i,k;j,k}+t^{\alpha \beta}_{k,i;k,j})\bigr] h_{ij}+$$

$$+\sum\limits_{i<j\atop {k<l}}(t^{\alpha}_{ij;kl}+t^{\beta}_{ij;kl})
\lbrack (ki|lj)-(kj|li)\rbrack +
\sum\limits_{i,j,k,l=1}^mt^{\alpha
\beta}_{i,j;k,l}(ki|lj)\biggr]\eqno(74)$$

where

$$\bar t^{\sigma}_{ij;kl}=\cases{+t^{\sigma}_{ij;kl} &if $i<j$ and $k<l$\cr
                                 -t^{\sigma}_{ij,lk} &if $i<j$ and $k>l$\cr
                                 -t^{\sigma}_{ji,kl} &if $i>j$ and $k<l$\cr
                                 +t^{\sigma}_{ji;lk} &if $i>j$ and $k>l$\cr
                                  \quad 0           &if $i=j$ or $k=l$\cr}
                                 \atop {(\sigma=\alpha ,\beta)}\eqno(75)$$

The property (5) of the exterior approximation for ${\cal W}_{2m,p,2}$
given by the convex set ${\cal V}_{2m,p,2}$ (see Sec.3) may be tested
numerically in the following rather obvious manner. With the aid of CI or
CASSCF method one can obtained $p$-electron wavefunction expansion over
determinant basis set and then contract the pure p-electron state to get
spin components of the 2-density operator.  Calculation of the lowest
eigenvalues of six 2-electron operators $t_2^{\alpha},
t_2^{\beta},t_2^{\alpha \beta},d_2^{\alpha},d_2^{\beta},$ and $d_2^{\alpha
\beta}$ with subsequent use of  Eq.(51b) to get $\varepsilon(t_2)$ makes
the problem of energy $E(t_2,\varepsilon(t_2))$ (see Eq.(22)) evaluation
at the border of ${\cal V}_{2m,p,2}$ very simple. It is pertinent to note
that we can take either the normalized identity matrix as a central point
of the set ${\cal V}_{2m,p,2}$ or its ``spin-adapted analogue'':

$$w_2=\frac{{p_{\alpha}\choose 2}}{{m\choose 2}{p\choose 2}}
\sum\limits_{i<j}|ij,\emptyset\rangle\langle ij,\emptyset |+
\frac{{p_{\beta}\choose 2}}{{m\choose 2}{p\choose 2}}
\sum\limits_{i<j}|\emptyset,ij\rangle\langle \emptyset ,ij |+
\frac{p_{\alpha}p_{\beta}}{m^2{p\choose 2}}
\sum\limits_{i,j}|i,j\rangle\langle i,j|\eqno(76)$$

For energy evaluation at the border points of ${\cal V}_{2m,p,2}$ the
choice of central point is of no consequence. At the same time for
optimization purpose central point defined by Eq.(76) should be used to
guarantee that at each step of energy minimization current 2-density
matrix has block-diagonal form (66). Note also that selection of the
central point (76) requires slight modification of the expression for the
distance function $\varepsilon (t_2)$ of the convex body ${\cal
V}_{2m,p,2}$.

Algorithms for contracting CI expansions to get the corresponding
2-density operator components $t_2^{\alpha},t_2^{\beta}$,and $t_2^{\alpha
\beta}$ and subsequent transformation of
these components with the aid of the operator $A(2m,p,2)$ were developed.
Using these algorithms in parallel with GAMESS program set \cite {GAMESS},
we performed calculations of CI wavefunctions for ground and excited
states of small atomic and molecular systems to get for each pure state
the distance function $\varepsilon(t_2)$ value to estimate its proximity
to the unit and to compare the energies at the border of ${\cal
V}_{2m,p,2}$ with the CI energies.  The detailed scheme of the algorithm
used can be described as follows.

(1) Calculate CI wavefunction $\Psi $ in determinant basis set;

(2) Contract pure state $|\Psi \rangle \langle \Psi |$ to get
spin components of 2-density matrix $t_2(\Psi );$

(3) Calculate spin components of the matrix $A(2m,p,2)t_2(\Psi );$

(4) Evaluate lowest eigenvalues of six 2-electron matrices
$t_2^{\sigma }(\Psi)$, $d_2^{\sigma }(\Psi )$ where
$\sigma =\alpha,\beta,\alpha \beta $,
and calculate the distance function value $\varepsilon(t_2(\Psi ));$

(5) Calculate energy of the central point used;

(6) Calculate  energy value at the point
$w_2 + \varepsilon(t_2(\Psi))[t_2-w_2]\in \partial {\cal V}_{2m,p,2}.$

Note that the final distance function and energy values are of no primary
interest  because they should be close to the unit and the CI energy
value, correspondingly, due to the property (5) of the convex set ${\cal
V}_{2m,p,2}$. From Eq.(51b) it follows that the
actual distance to the border of ${\cal V}_{2m,p,2}$ is determined as the
minimum of distances from the central point to the border of ${\cal
E}_{2m,2}$ and the border of $A^{-1}(2m,p,2){\cal E}_{2m,2}$.  Of certain
interest is which of these two distances determines the final one for
concrete pure CI state.

The results of trial calculations are listed in Tables 1 and 2. In atomic
calculations cc\_ pvDZ basis set of Dunning \cite {Dunning} was employed.
For lithium, berrilium, and boron FCI calculations were carried out,
whereas for carbon and nitrogen $1s$ AO, and for oxygen $1s,2s$ AOs were
excluded from the active spaces to keep the sizes of CI expansions
reasonable for running GAMESS on PC with 166 MHz Intel processor.
In the case of molecules cc\_pvDZ
basis and FCI were used for calculation of LiH, and 6-31G Gaussian basis
set \cite {Ditch, Hehre-1} with frozen $1s$ AO for CH$_2$, and $1s,2s$ AOs
for NH$_2$,H$_2$O, and NH$_3$ was employed.

As seen from Tables 1 and 2, the exterior approximation for pure CI states
given by the convex set ${\cal V}_{2m,p,2}$ is very good as it was
expected.  It is to be noted that the contraction procedure may lead to
roundoff errors in matrix elements of 2-density matrices, especially when
CI expansion is large. As a result, diagonalization of 2-density matrix
may give small negative values for its lowest eigenvalues in spite of the
fact that the matrix $t_2(\Psi )$ for any $\Psi $ is manifestly
nonnegative.  For this reason in Tables 1 and 2 the absolute values of
$\varepsilon (t_2)$ deviation from the unit are given. Of interest is also
the fact that for atoms and high symmetry linear molecules the distance
from the central point to the border of ${\cal V}_{2m,p,2}$ is determined
by the lowest eigenvalue of the relevant density operator whereas for low
symmetry molecules this distance is determined by the lowest eigenvalue of
the operator $A(2m,p,2)t_2$.

\newpage
\centerline{\bf 6. Algorithms for Direct Determination of 2-Density Matrix}
\bigbreak

The energy expression (74) may be essentially simplified by turning to a
new basis set in the 2-electron section of the Fock space.  Indeed, let us
consider the basis set of eigenvectors of three matrices

$$H^{\alpha}_{ij;kl}={{\partial E(t_2)}\over {\partial t_{ij;kl}^{\alpha}}}
\eqno(77a)$$
$$H^{\beta}_{ij;kl}={{\partial E(t_2)}\over {\partial t_{ij;kl}^{\beta}}}
\eqno(77b)$$
$$H^{\alpha \beta}_{i,j;k,l}={{\partial E(t_2)}\over {\partial
t_{i,j;k,l}^{\alpha \beta}}}\eqno(77c)$$

that are just (up to fixed prefactor) spin blocks of the reduced
Hamiltonian $H_{p\to 2}$. In this ``energy'' basis arbitrary 2-density
operator is of  the form

$$t_2=
\sum\limits _{r,s=1}^{m\choose 2}
\lambda_{rs}^{\alpha}|\Phi_r^{\alpha}\rangle \langle \Phi_s^{\alpha}|+
\sum\limits _{r,s=1}^{m\choose 2}
\lambda_{rs}^{\beta}|\Phi_r^{\beta}\rangle \langle \Phi_s^{\beta}|+
\sum\limits _{r,s=1}^{m^2}
\lambda_{rs}^{\alpha \beta}|\Phi_r^{\alpha \beta}\rangle \langle
\Phi_s^{\alpha \beta}|\eqno(78)$$

and the electronic  energy expression may be  written

$$E(t_2)=
\sum\limits_{r=1}^{m\choose 2}\lambda _{rr}^{\alpha}\epsilon  _r^{\alpha}+
\sum\limits_{r=1}^{m\choose 2}\lambda _{rr}^{\beta}\epsilon _r^{\beta}+
\sum\limits_{r=1}^{m^2}\lambda _{rr}^{\alpha \beta}\epsilon _r^{\alpha \beta}
\eqno(79)$$

Here $\{\epsilon _r^{\alpha}, |\Phi _r^{\alpha}\rangle \}$,
$\{\epsilon _r^{\beta}, |\Phi _r^{\beta}\rangle \}$, and
$\{\epsilon _r^{\alpha \beta}, |\Phi _r^{\alpha \beta}\rangle \}$
are the eigenvalues and eigenvectors of the matrices (77a), (77b), and
(77c), correspondingly.

Since in the energy representation the electronic energy is a linear
functional involving only the diagonal elements of 2-density matrix it
seems reasonable to try the well-known Box method
\cite {Box} for energy optimization. This method
starts with generating at least
$\kappa={2m\choose 2}+1$ affine independent diagonal matrices
$\lambda^{(1)},\lambda^{(2)},\ldots , \lambda^{(\kappa )}$ with nonegative
entries such that for each diagonal matrix $\lambda^{(i)}$ from this set
the following Approximate Representability condition ($AR$-condition) is
satisfied:

$(AR)$ {\sl There exists (at least one) symmetric matrix} $\Lambda^{(i)} $
{\sl with zero diagonal entries such that} $$U\lambda^{(i)} U^{\dagger} +
U\Lambda^{(i)} U^{\dagger}\in {\cal V}_{2m,p,2}\eqno(80)$$ {\sl where} $U$
{\sl is the unitary transformation from the 2-electron determinant basis
to the energy basis.}

It is pertinent to note here that the convex set ${\cal V}_{2m,p,2}$ being
invariant with respect to the orbital unitary transformations is actually
changed under 2-electron transformations.

Then we should order the initial vertices in the energy increasing order:

$$E(\lambda^{(1)} ) \le E(\lambda^{(2)}) \le \ldots
\le E(\lambda^{(\kappa )})\eqno(81)$$

and calculate the weight center of the first $\kappa -1$ vertices

$$\bar \lambda={1\over {\kappa -1}} \sum\limits_{i=1}^{\kappa -1}
\lambda^{(i)}\eqno(82)$$

This weight center obviously satisfies the $AR$-condition and is used
in the Box method for reflection of the "worst" vertex $\lambda^{(\kappa
)}$:

$$\lambda (\epsilon)=\bar \lambda +\epsilon
(\bar \lambda - {\lambda}^{(\kappa )})\eqno(83)$$

where $\epsilon$ is a positive reflection coefficient. It is clear that
$E(\bar \lambda) \le E({\lambda }^{(\kappa )})$ and that the energy value
may only decrease when moving along the vector $\bar \lambda -
{\lambda}^{(\kappa )}$.  The optimal value ${\epsilon}_*$ of the
reflection coefficient should satisfy the following condition:

$(\overline {AR})$ {\sl There exists (at least one) symmetric matrix}
$\Lambda $  {\sl with zero diagonal entries such that}
$$U\lambda ({\epsilon }_* )U^{\dagger} + U\Lambda U^{\dagger}
\in \partial {\cal V}_{2m,p,2}\eqno(84)$$
{\sl where} $U$ {\sl is the unitary transformation from the
2-electron determinant basis to the energy basis.}

The next step of the Box algorithm consists in replacing the "worst"
vertex ${\lambda }^{(\kappa )}$ by the calculated optimal vertex $\lambda
({\epsilon }_*)$ with subsequent reordering the new set in the energy
increasing order. Then the new weight center should be constructed and new
optimal vertex should be determined.  This process is repeated till the
energy value stabilization and the final weight center is taken as the
desired solution.

The crucial point of the algorithm described is the calculation of the
optimal vertex satisfying the $\overline {AR}$-condition. Let us suppose
that the current 2-density matrix is of the form

$$t_2=t_{\lambda } + U\Lambda U^{\dagger} \eqno(85)$$

where $diag(U^{\dagger}t_{\lambda}U)=\lambda $, and ${\Lambda }$ is some
symmetric matrix with zero diagonal entries.  The distance of matrix (85)
from some symmetric nonnegative 2-density matrix $X^2$ (in the standard
trace metric) is given by the function

$$f^{(1)}_{\lambda}(\Lambda,X)=Tr\big (t_{\lambda }+U\Lambda U^{\dagger}
-X^2\big )^{\dagger }\big (t_{\lambda }+U\Lambda U^{\dagger}-X^2\big
)\eqno(86a)$$

whereas  the distance of this matrix from some matrix $A^{-1}(Y^2)$ is

$$f^{(2)}_{\lambda}(\Lambda,Y)=Tr\big (t_{\lambda }+U\Lambda U^{\dagger}
-A^{-1}(Y^2)\big )^{\dagger } \big (t_{\lambda }+
U\Lambda U^{\dagger}-A^{-1}(Y^2)\big )\eqno(86b)$$

Here $X$ and $Y$ are symmetric matrices of parameters that have the same
block-daigonal structure as 2-density matrix (66) and satisfy the
restrictions on block trace values given be Eqs.(67a)-(67b) and
Eqs.(69a)-(69b), correspondingly.

The function

$$f_{\lambda}(\Lambda,X,Y)=f^{(1)}_{\lambda}(\Lambda,X)+
f^{(2)}_{\lambda}(\Lambda,Y)\eqno(87)$$

can be considered as as a kind of characteristic function of the set
${\cal V}_{2m,p,2}$ of all approximately representable 2-density matrices.
Indeed, if $t_2\in {\cal V}_{2m,p,2}$ with $diag(U^{\dagger}t_2U)=\lambda$
then

$$f_{\lambda}(0,t_2^{1\over 2},[A(2m,p,2)t_2]^{1\over 2})=0$$

On the other hand, if for fixed $t_{\lambda}$ there exist matrices $\Lambda$,
$X$, and $Y$, such that

$$f_{\lambda}(\Lambda,X,Y)=0$$

then, due to the nondegeneracy of the scalar product used,
$t_2=t_{\lambda}+ U\Lambda U^{\dagger}=X^2=A^{-1}(2m,p,2)(Y^2)$ belongs to
${\cal V}_{2m,p,2}$.

Now we can formulate the $(AR)$ condition in the form admitting
numerical verification:

$(AR)'$ {\sl If for a fixed diagonal 2-density matrix} $\lambda$ {\sl
there exist symmetric matrix} $\Lambda$ {\sl with zero diagonal entries
and symmetric matrices of parameters} $X$ {\sl and} $Y$ {\sl such that}
$$f_{\lambda}(\Lambda,X,Y)=0$$ {\sl }then $$U\lambda U^{\dagger} +
U\Lambda U^{\dagger}\in {\cal V}_{2m,p,2}$$

To calculate the optimal vertex $\lambda ({\epsilon }_*)$ it is necessary
to analyze the nonnegative scalar function

$$f(\epsilon) = \min\limits_{\Lambda,X,Y} f_{\lambda (\epsilon)}
(\Lambda,X,Y)\eqno(88)$$

that is strictly positive for $\epsilon > {\epsilon }_*$ and accepts zero
values in the interval $[0,{\epsilon }_*]$.

Note that the initial set of vertices in the Box method should be
carefully selected to ensure their affine independence because in the
opposite case, due to linear character of this method, the search of
optimal vertex will be performed only within a face of the polyhedron
spanned by the initial vertices.

The method described being conceptually very simple is not easy for
practical implementation. Its expected main drawbacks are:

(1) It requires often transformations from 2-electron determinant
basis to the energy one and back;

(2) Optimization algorithm include both differentiable and non-
differentiable parts;

(3) Minimization of function (87) with respect to $X$ and $Y$ should be
performed with very high accuracy (about $10^{-20}$ in gradient norm);

(4) Function (87) has excessive number of parameters that results in
degenerate Hessians and low convergence of minimization process.

There exists another approach based on the energy expression modification.
Since operator $A(n,p,q)$ is symmetric and invertable (see Sec.2 and
Appendix C), we can write down the following equality:

$$Tr[H_{p\to 2}t_2]=Tr[(A^{-1}(2m,p,2)H_{p\to 2})A(2m,p,2)t_2]$$

Introducing two symmetric matrices of independent variables we can rewrite
the energy expression (22) in the form

$$E(X,Y)= \frac{{p\choose 2}}{2}\biggl (Tr[H_{p\to 2}X^2] +
Tr[(A^{-1}(2m,p,2)H_{p\to 2})Y^2]\biggr )\eqno(89)$$

The corresponding optimization problem

$$\cases {\min\limits_{X,Y}E(X,Y)\cr
           A(2m,p,2)X^2=Y^2\cr
           Tr(X^2)=1\cr
           Tr(Y^2)=1\cr}\eqno(90)$$

may be solved by constrained optimization technique (see, e.g. \cite
{Powell}).  This approach, in contrast to the first one, does not require
2-electron unitary transformations and both the energy expression (89) and
the restrictions on variables are differentiable. Its main drawback is
shared by all methods based on algorithms involving Lagrange multipliers
and penalty functions: in theory infinite number of intermediate
unconstrained optimizations is required to reach the solution. It is to be
noted as well that the number of variables in Eq.(90) can be essentially
reduced if we take into account that matrices $X$ and $Y$ should be of
block-diagonal structure for states with fixed total spin projection. If
block-diagonal structure of $X$ and $Y$ is explicitly accounted, it is
reasonable to replace trace resrictions for $X^2$ and $Y^2$ by six trace
restrictions for the corresponding spin blocks (see Eqs.(67a)-(67b) and
(69a)-(69b)).

\bigbreak
\centerline{\bf 6. Conclusion}
\bigbreak
The exterior approximation for the set of all representable density
operators of arbitrary order described in this work is expected to be
useful for direct 2-density matrix determinantion. However, only on the
base of concrete calculations it will be possible to estimate the
practical importance of the approximation obtained.

Two algorithms for such calculations are developed.  In contrast to the
standard approaches where the electronic energy domain turns out to be one
of classic analytic manifolds (unit sphere, orthogonal group and its
quotients, etc) that can be easily parametrized by, say, elements of
relevant tangent spaces \cite{Panin 1}, the convex set ${\cal V}_{2m,p,2}$
is of much more complicated nature. The only thing we can try do at present
is to cover
${\cal V}_{2m,p,2}$ using excessive set of parameters.  Energy
optimization on ${\cal V}_{2m,p,2}$ may be a very complicated
computational problem but if efficient optimization scheme is developed,
there may be opened a way to FCI quality calculation of fairly extensive
molecular systems.

\newpage
\centerline{\bf Appendix A.}
\bigbreak
Let $N=\{1,2,\ldots,n\}$ be the spin-orbital index set. On the set
${\cal P}(N)$ of all subsets of $N$ let us consider the operation

$$R\Delta S=(R\cup S)\backslash (R\cap S)\eqno(A.1)$$

where $R,S\in {\cal P}(N)$. This operation endows ${\cal P}(N)$ with
Abelian group structure with empty set as its unit. Each element of this
group is of order 2 ($R\Delta R=\emptyset $). The mapping

$${\varphi}_{\Delta}:K\to {\Delta}_K\eqno(A.2)$$

where $K\subset N$ and

$${\Delta}_K=\Delta_{k\in K}\{1,2,\ldots ,k\}\eqno(A.3)$$

is a group homomorphism. Indeed, ${\Delta}_{\emptyset}=\emptyset$ and
$({\Delta}_K)\Delta ({\Delta}_L)={\Delta}_{K\Delta L}$.

In particular,
$${\Delta}_{\{k\}}=\{1,2,\ldots,k\}\eqno(A.4a)$$
$${\Delta}_{\{k,l\}}=\cases{ \{k+1,\ldots,l\} &if k$<$l \cr
                          \{l+1,\ldots,k\} &if k$>$l \cr}\eqno(A.4b)$$
$$\Delta_N=\cases {\{2,4,\ldots\} &if n {is even} \cr
                          \{1,3,\ldots\} &if n {is odd}\cr}\eqno(A.4c)$$
In general case, for  $K=k_1<k_2<\ldots<k_s$
$${\Delta}_K=\cases{
\bigcup\limits_{i=1}^{[{s\over 2}]}\{k_{2i-1}+1,\ldots ,k_{2i}\}
&if s is even\cr
\bigcup\limits_{i=0}^{[{s\over 2}]}\{k_{2i}+1,\ldots ,k_{2i+1}\}
&if s is odd\cr}
                    \eqno(A.5)$$

Directly from the definition of operation $\Delta$ the following relations
important for phase prefactors evaluation may be obtained

$$|K\cap R|+|K\cap S|\equiv |K\cap (R\Delta S)|{\ } (mod{\ } 2)\eqno(A.6)$$
$$|K\cap \Delta_K|=\lbrack {{|K|+1}\over 2}\rbrack\eqno(A.7)$$

\newpage
\centerline{\bf Appendix B.}
\bigbreak
From Eq.(30) it is easy to get matrix representation of the operator
$A^{-1}(n,p,q)$ with respect to the basis $\{e^{IJ}_K\}$ for $p+q=n$.
Confining ourselves to the diagonal case $I=J=\emptyset $, we obtain

$$A^{-1}_{K'K}(n,p,q)=\cases{{1\over {p\choose q}},
&if $K'\subset N\backslash K$\cr
0,&if $K'\not \subset N\backslash K$\cr}
\eqno(B.1)$$

Let us suppose that
$$A_{KK'}(n,p,q)=(-1)^{|K\cap K'|}f(p,q,|K\cap K'|)\eqno(B.2)$$

$A(n,p,q)$ matrix elements are determined by the equations:

$$\sum\limits_{K'}^{(q)}A_{KK'}(n,p,q)A^{-1}_{K'K''}(n,p,q)=$$

$$={1\over {p\choose q}}\sum\limits_{K'\subset N\backslash
K''}^{(q)}(-1)^{|K\cap K'|}
f(p,q,|K\cap K'|)=\delta_{K,K''}\eqno(B.3)$$

After simple set-theoretical manipulations we come to the following
combinatorial equations

$$\sum\limits_{r=0}^t(-1)^r{{p-t}\choose {q-r}}{t\choose
r}f(p,q,r)={p\choose q}
\delta _{t,0}\eqno(B.4)$$

where $t=|K\backslash (K\cap K'')|$ and $r=|K\cap K'|$.
From this equation it is easy to get reccurently the expression for
$A_{KK'}(n,p,q)$:

$$A_{KK'}(n,p,q)=(-1)^{K\cap K'|}{{{p+|K\cap K'|-q-1}\choose {|K\cap K'|}}
\over {{p\choose q}{q\choose {|K\cap K'|}}}}\eqno(B.5)$$

Note that Eq.(B.4) can be rewritten in the form

$$\sum\limits_{r=0}(-1)^r{t\choose r}{{p-t}\choose {q-r}}
{{{p+r-q-1}\choose r}\over {q\choose r}}=
(-1)^q{p\choose q}
{{t-1}\choose q}\eqno(B.6)$$

and proved by induction for $p\ge q \ge 1$  and $t=0,1,\ldots ,p$.

\newpage
\centerline{\bf Appendix C.}
\bigbreak
Let us reduce the problem of operator $A(n,p,q)$ inversion to
combinatorial equations. The matrix elements of the operator $A(n,p,q)$
with respect to the operator basis $(e^{IJ}_{K})$ are given by Eq.(43).
Since this matrix is block-diagonal, and all its blocks are similar in
their structure, we can confine ourselves to the block with
$I=J=\emptyset$.

Let us consider the system of linear equations

$$\sum\limits_{K'\subset N}^{(q)}A_{KK'}^{\emptyset \emptyset}(n,p,q)
X_{K'K''}(n,p,q)=\delta_{|K\cap K''|,q}\eqno(C.1)$$

for determining the inverse matrix (block). Simple combinatorial arguments
together with the additional assumption that $X_{K'K''}=f(n,p,q;|K'\cap
K''|)$ (that is $X_{K'K''}$ depends not on subsets $K',K''$ but only on
the number of elements in their intersection) allow us to rewrite the
system (C.1) as the following system of combinatorial equations

$$\sum\limits_{r_1,r_2,t}(-1)^{r_1}
{{p+r_1-q-1\choose r_1}\over {q\choose r_1}}{q-u\choose r_1-t}
{q-u\choose r_2-t}{u\choose t}{n-2q+u\choose q-(r_1+r_2-t)}f(n,p,q;r_2)$$
$$={n-p\choose q}\delta_{u,q}\eqno(C.2)$$

where $u=|K\cap K''|=0,1,\ldots,q$, $r_1=|K\cap K'|=0,1,\ldots,q$,
$r_2=|K'\cap K''|=0,1,\ldots,q$, and $t=|K\cap K'\cap K''|=0,1,\ldots,u$.
Thorough analysis of this system leads to the conclusion that

$$f(n,p,q;r_2)=(-1)^{r_2}{{n-p-q+r_2-1\choose r_2}\over {{p\choose q}
{q\choose r_2}}}\eqno(C.3)$$

and, as a result,  the operator $A(n,p,q)$ is
invertible and  the matrix elements of the inverse matrix  are

$$ (A^{-1})_{KK'}^{IJ}(n,p,q)= (-1)^{\alpha_{IJ}+|K\cap K'|}{{n-p\choose
q}\over{{n-p-s\choose {q-s}}{p\choose q}}} {{n-p-q+|K\cap K'|-1\choose
{|K\cap K'|} }\over {{q-s\choose |K\cap K'|}}}\eqno(C.4)$$

where $|I|=|J|=s,I\cap J=\emptyset,K,K'\subset N\backslash (I\cup
J),|K|=|K'|=q-s $.  The last equation implies, in particular, that

$$A^{-1}(n,p,q)=A(n,n-p,q)\eqno(C.5)$$

\newpage
\centerline{\bf Appendix D.}
\bigbreak

{\bf Lemma}.
$$A(n,p,q){\cal W}_{n,p,q}\subset {\cal E}_{n,q}\eqno(D.1)$$
{\bf Proof}. It is sufficient to show that $A(n,p,q){q!\over {p!}}c^{p-q}
|\Psi\rangle\langle\Psi|\in {\cal E}_{n,q}$
for arbitrary  $\Psi\in {\cal F}_{n,p}$. Expanding $\Psi$ over p-electron
determinant basis set
$$\Psi=\sum\limits_{R\subset N}^{(p)}C_R|R\rangle$$
and turning to the operator basis $(e_L^{IJ})$ in
${\cal F}_{n,p}\otimes {\cal F}^*_{n,p} $ we obtain
$$\frac{q!}{p!}c^{p-q}|\Psi\rangle\langle\Psi|=
\sum\limits_{s=0}^q\sum\limits_{I,J}^{(s)}
\sum\limits_{L\subset N\backslash(I\cup J)}^{(p-s)}
(-1)^{|(I\cup J)\cap \Delta_L|}C_{I\cup L}C^*_{J\cup L}\frac{{p-s\choose q-s}}
{{p\choose q}}w^{IJ}_{p\downarrow q}(L)\eqno(D.2)$$
where symmetric combinations  $w^{IJ}_{p\downarrow q}(L)$
of q-electron  basis operators
are given by Eq.(32).  Simple set-theoretical and combinatirial manipulations
lead to the equality
$$A(n,p,q)w^{IJ}_{p\downarrow q}(L)=
(-1)^{\alpha_{IJ}}\frac{{p\choose q}}{{p-s\choose q-s}{n-p\choose q}}
\sum\limits_{K\subset N\backslash (I\cup J\cup L)}^{(q-s)}e_K^{IJ}\eqno(D.3)$$
Turning back  to q-electron determinant generators and carefully
handling the phase prefactors arising, we arrive at
$$A(n,p,q)\frac{q!}{p!}c^{p-q}|\Psi\rangle\langle \Psi|=
\frac{1}{{n-p\choose q}}\sum\limits_{K\subset N}^{(n-p-q)}\lbrack
\sum\limits_{R\subset N\backslash K}^{(p)}(-1)^{|R\cap \Delta_K|}
C_R|N\backslash (R\cup K)\rangle\rbrack\times $$
$$\lbrack\sum\limits_{S\subset {N\backslash K}}^{(p)}
(-1)^{|S\cap \Delta_K|}C_S^*\langle N\backslash (S\cup K)|\rbrack=
\frac{1}{{n-p\choose q}}\sum\limits_{K\subset N}^{(n-p-q)}
|\Phi_K\rangle\langle\Phi_K|\eqno(D.4)$$
The operator on the right-hand side of this equation is obviously
positive definite. To complete the proof, we should check the
normalization property that  can be easily established by
direct calculation $\blacksquare$

Let us recast the last equation in a more convenient form that makes
its structure more transparent. We have
$$A(n,p,q)\frac{q!}{p!}c^{p-q}|\Psi\rangle\langle \Psi|=
\frac{1}{{n-p\choose q}}\sum\limits_{Z\subset N}^{(p+q)}d_q(Z)\eqno(D.5)$$
where
$$d_q(Z)=|\Psi_Z\rangle \langle\Psi_Z|\eqno(D.6)$$
and
$$|\Psi_Z\rangle =\sum\limits_{R\subset Z}^{(q)}(-1)^{|(Z\backslash R)
\cap \Delta_{N\backslash Z}|}C_{Z\backslash R}|R\rangle \eqno(D.7)$$
Eqs.(D.5)-(D.7) may serve as  a base for further analysis of both pure and
ensemble representability problems. For example, at first glance the necessary
condition  of the ensemble representability of $t_q$
\bigbreak
(ER) {\sl There exists an expansion of $d_q=A(n,p,q)t_q$ of the type of
Eq.(D.5) such that for
any  $Z\subset N, |Z|=p+q$ the operator $d_q(Z)$ is nonnegative}
\bigbreak
looks much stronger than simple positive semidefinitness  of $d_q$.

The necessary and sufficient conditions of pure representability may be
formulated with the aid of Eqs.(D.5)-(D.7). Let us introduce the set
$$B_{n,p,q}=\{(Z,R)\subset N\times N:|Z|=p+q \& |R|=q \& R\subset Z\}
\eqno(D.8)$$
and the equivalence relation on this set
$$ (Z,R) \sim (Z',R') \Leftrightarrow Z\backslash R=Z'\backslash R'
\eqno(D.9)$$
The set of all equivalence classes $\bar B_{n,p,q}$ contains $n\choose p$
elements and in each equivalence class $\overline {(Z,R)}$ there are
$n-p\choose q$ elements.

{\bf Theorem}. {\sl $q$-electron operator $t_q$ is representable by pure
$p$-electron state if and only if its image $d_q$ with respect to
$A(n,p,q)$ satisfies the following conditions:

(i) There exists expansion
$$d_q=\frac{1}{{n-p\choose q}}\sum\limits_{Z\subset N}^{(p+q)}d_q(Z)
\eqno(D.10)$$
such that for each $Z\subset N$ $d_q(Z)$ is either the null operator or
corresponds to (unnormalized) pure $q$-electron state $|\Psi_Z\rangle$;

(ii) The mapping
$$ (Z,R) \to (-1)^{|(Z\backslash R)\cap \Delta_{N\backslash Z}|}
\langle \Psi_Z|R\rangle \eqno(D.11)$$
is constant on the equivalence classes $\overline {(Z,R)}$.}

If the conditions of this theorem are fullfilled then (up to
normalization) the required pure state may be presented as
$$ |\Psi\rangle=\sum\limits_{\overline {(Z,R)}}(-1)^
{|(Z\backslash R)\cap \Delta_{N\backslash Z}|}\langle\Psi_Z|R\rangle
|Z\backslash R\rangle \eqno(D.12)$$
where the sum runs over equivalence classes.
Unfortunately, condition (i) of the above theorem is depressingly
non-constructive.
Only for $p+q=n$ conditions (i) and (ii)
are trivial and the following corollary of the main theorem may be
formulated.

{\bf Corollary}. {\sl For $p+q=n$ $q$-electron operator $t_q$
is representable by pure
$p$-electron state if and only if its image $d_q$ with respect to
$A(n,p,q)$ corresponds to pure $q$-electron state.}

\newpage
\centerline{\bf Appendix E.}
\bigbreak
Let us show how by careful choice of the phase prefactors in Eq.(43) it is
possible to ensure the unitary invariance of the convex set
${\cal V}_{n,p,1}$.
We have
$$A(n,p,1)e^{\emptyset \emptyset}_k=
(-1)^{{\alpha}_{\emptyset \emptyset}}
{1\over {n-p}}\biggl[ \sum\limits_{k'}
e^{\emptyset \emptyset}_{k'} -
pe^{\emptyset \emptyset}_{k}\biggr]\eqno(E.1)$$
$$A(n,p,1)e^{ij}_{\emptyset }=
(-1)^{{\alpha}_{ij}}{p\over {n-p}}e^{ij}_{\emptyset}\eqno(E.2)$$

For arbitrary unitary operator $u\in U({\cal F}_{n,1})$
$$A(n,p,1)u\otimes u^{\dag}
e^{\emptyset \emptyset}_{k} =
(-1)^{{\alpha}_{\emptyset \emptyset}}{1\over {n-p}}
\sum\limits_{k_1}e^{\emptyset \emptyset}_{k_1}
(1-pu_{k_1k}u^*_{k_1k})+$$
$$+{p\over {n-p}}
\sum\limits_{k_1,k_2\atop {(k_1\ne k_2)}} (-1)^{{\alpha}_{k_1k_2}}
e^{k_1 k_2}_{\emptyset}
u_{k_1k}u^*_{k_2k}\eqno(E.3)$$
$$u\otimes u^{\dag}A(n,p,1)
e^{\emptyset \emptyset}_{k} =
(-1)^{{\alpha}_{\emptyset \emptyset}}{1\over {n-p}}
\sum\limits_{k_1}e^{\emptyset \emptyset}_{k_1}
(1-pu_{k_1k}u^*_{k_1k})-$$
$$-(-1)^{{\alpha}_{\emptyset \emptyset}}{p\over {n-p}}
\sum\limits_{k_1,k_2\atop {(k_1\ne k_2)}}e^{k_1 k_2}_{\emptyset}
u_{k_1k}u^*_{k_2k}\eqno(E.4)$$
$$A(n,p,1)u\otimes u^{\dag}
e^{ij}_{\emptyset} =
-(-1)^{{\alpha}_{\emptyset \emptyset}}{p\over {n-p}}
\sum\limits_{k_1}e^{\emptyset \emptyset }_{k_1}
u_{k_1i}u^*_{k_1j}+$$
$$+{p\over {n-p}}
\sum\limits_{k_1,k_2\atop {(k_1\ne k_2)}}(-1)^{{\alpha}_{k_1k_2}}
e^{k_1 k_2}_{\emptyset}
u_{k_1i}u^*_{k_2j}\eqno(E.5)$$
$$u\otimes u^{\dag}A(n,p,1)
e^{ij}_{\emptyset} =
(-1)^{{\alpha}_{ij}}{p\over {n-p}}
\sum\limits_{k_1}e^{\emptyset \emptyset}_{k_1}
u_{k_1i}u^*_{k_1j}+$$
$$+(-1)^{{\alpha}_{ij}}{p\over {n-p}}
\sum\limits_{k_1,k_2\atop {(k_1\ne k_2)}}e^{k_1 k_2}_{\emptyset}
u_{k_1i}u^*_{k_2j}\eqno(E.6)$$
From these equalities it is readily follows that
to ensure the unitary invariance of $A(n,p,1)$ it is sufficient to require
that either
$${\alpha}_{\emptyset \emptyset}\equiv 0(mod{\ }2) \Rightarrow
 {\alpha}_{ij}\equiv 1(mod{\ }2)\eqno(E.7a)$$
 or
$${\alpha}_{\emptyset \emptyset}\equiv 1(mod{\ }2) \Rightarrow
 {\alpha}_{ij}\equiv 0(mod{\ }2)\eqno(E.7b)$$
 for every $i,j=1,\ldots,n {\ }(i\ne j)$.
Our choice ${\alpha}_{\emptyset \emptyset}=0$ and
${\alpha}_{ij}=|\{i,j\}\cap \Delta _{\{i,j\}}|$ for every $i\ne j$
corresponds to the implication (E.7a).

\newpage
\centerline{\bf Appendix F.}
The commutation relation (54) in coordinate form looks like
$$\sum\limits_{K\subset L}^{(q-s-1)}A_{K'K}^{IJ}(n,p,q-1)=
\sum\limits_{L'\supset K'}^{(q-s)}A^{IJ}_{L'L}(n,p,q)\eqno(F.1)$$
where $|L|=|L'|=q-s$, and $ L,L' \subset N\backslash (I\cup J)$.

Direct calculations show that
$$\sum\limits_{K\subset L}^{(q-s-1)}(-1)^{|K\cap K'|}
{{{p+|K\cap K'|-q}\choose {|K\cap K'|}}\over {{q-s-1}\choose {|K\cap K'|}}}=
(-1)^t{{{p+t-q}\choose t}\over {{q-s}\choose t}}{(p-q)(q-s)\over {p+t-q}}
\eqno(F.2)$$
and
$$\sum\limits_{L'\supset K'}^{(q-s)}(-1)^{|L\cap L'|}
{{{p+|L\cap L'|-q-1}\choose {|L\cap L'|}}\over {{q-s}\choose {|L\cap L'|}}}=
(-1)^t{{{p+t-q}\choose t}\over {{q-s}\choose t}}{(p-q)(n-p-q+1)\over {p+t-q}}
\eqno(F.3)$$
where $t=|K'\cap L|$.

Taking into account binomial prefactors in expression (43) leads readily to
Eq.(F.1) that is equivalent to the commutation relation (54).

It may be expected that there exist many reccurence relations of the
type of Eq.(F.1) involving matrix elements of the operator A(n,p,q).

\newpage
\centerline{\bf Appendix G.}
{\bf Lemma.}{\sl Vectors $w^{IJ}_{q\uparrow p}(K)$ defined by Eq.(33) are
lineary independent.}

{\bf Proof.} Let us consider the vector equation
$$\sum\limits_{K\subset N\backslash (I\cup J)}^{(q-s)}{\mu}^{IJ}_K
w^{IJ}_{q\uparrow p}(K)=0$$
Using definitions (32) and (33), we can rewrite this equation
in the form
$$\sum\limits_{L\subset N\backslash (I\cup J)}^{(p-s)}e^{IJ}_L
Tr[({\mu}^{IJ})^{\dag}w^{IJ}_{p\downarrow q}(L)] =0$$
that is equivalent to the system of ${n-2s\choose p-s}$ of scalar equations
$$Tr[({\mu}^{IJ})^{\dag}w^{IJ}_{p\downarrow q}(L)] =0,
L\subset N\backslash (I\cup J)$$
But the set of vectors $w^{IJ}_{p\downarrow q}(L)$ is
complete in the subspace of the operator space ${\cal F}_{n,q}\otimes
{\cal F}_{n,q}^*$ determined by disjoint index sets $I$ and $J$
(see \cite {Panin})$\blacksquare$

Following general scheme of Sec.4, we calculate
$$\frac{q!}{p!}c^{p-q}\frac{(n-p)!}{(n-q)!}u^{p-q}e^{IJ}_K=
\frac{{p-s\choose q-s}{n-q-s\choose p-q}}{{p\choose q}{n-q\choose p-q}}
\sum\limits_{K'\subset N\backslash (I\cup J)}^{(q-s)}e^{IJ}_{K'}
[G^{IJ}(n,p,q)]_{K'K},\eqno(G.1)$$
where
$$[G^{IJ}(n,p,q)]_{K'K}=\frac{{n-2q+|K'\cap K|\choose p-2q+s+|K'\cap K|}}
{{p-s\choose q-s}{n-q-s\choose p-q}}\eqno(G.2),$$
and $s=|I|=|J|$.
Note that concrete blocks $G^{IJ}(n,p,q)$  thus introduced differ from
blocks that could be constructed on the base of general definition (58)
by non-essential combinatorial prefactor. The full-size block-diagonal
matrix $G(n,p,q)$ is
$$G(n,p,q)=\bigoplus\limits_{s=0}^q\bigoplus\limits_{I,J}^{(s)}
\frac{{p-s\choose q-s}{n-q-s\choose p-q}}{{p\choose q}{n-q\choose p-q}}
G^{IJ}(n,p,q)\eqno(G.3)$$
Since for any fixed  $I,J$ all blocks $G^{IJ}(n,p,q)$ are of similar
structure we can confine ourselves to analysis of the ``diagonal'' case
$I=J=\emptyset$ that corresponds to $s=0$ in Eqs.(G.1)-(G.3).

The problem of $G^{\emptyset \emptyset}(n,p,q)$ inversion can be
reformulated as a combinatorial problem if we suppose that
$[G^{\emptyset \emptyset}(n,p,q)]^{-1}_{KK''}=f(n,p,q;|K\cap K''|)$.
It turns out, however, that we do not need $[G(n,p,q)]^{-1}$
in itself but only the  sums of the type
$$b(n,p,q;|K'\cap L|)=\sum\limits_{K\subset L}[G^{IJ}(n,p,q)]^{-1}_{K'K}
\eqno(G.4)$$
as can be easily demonstrated by calculation of the expansion operator action
on $[G(n,p,q)]^{-1}t_q$.
After rather complicated combinatorial manipulations we arrive at
the linear system
$$\sum\limits_{r_1,r_2,t}{n-2q+r_1\choose p-2q+r_1}
{n-p-q+u\choose q-(r_1+r_2-t)}
{q-u\choose r_1-t}{p-u\choose r_2-t}{u\choose t}b(n,p,q;r2)=$$
$${p\choose q}{n-q\choose n-p}{\delta}_{uq},\eqno(G.5)$$
where $u=|K'\cap L|$.

Its solution for all $n,p,q$ $(q\le p\le n)$  is
$$b(n,p,q;r_2)=(-1)^{q+r_2}{p\choose q}\frac{{p-r_2-1\choose q-r_2}}
{{n-p\choose q-r_2}}.\eqno(G.6)$$
and our final formula necessary for explicit construction of section (60) is
$$\sum\limits_{K\subset L}[G^{IJ}(n,p,q)]^{-1}_{K'K}=
(-1)^{q+s+|K'\cap L|}{p-s\choose q-s}\frac{{p-s-|K'\cap
L|-1\choose q-s-|K'\cap L|}}{{n-p-s\choose q-s-|K'\cap L|}}.\eqno(G.7)$$

\bigbreak
\bigbreak
\bigbreak

\centerline{\bf Acknowledgment}
\bigbreak
We gratefully acknowledge the Russian Foundation for Basic Research
(Grant 00-03-32943a) and Ministry of Education of RF
(Grant E00-5.0-62) for financial support of the present work

\newpage
\begin{table}[ht]
\caption{Results of simple calculations testing properties
of the convex set ${\cal V}_{2m,p,2}: $ atoms}
\vspace{10mm}
\begin{tabular}{|l|c|c|c|c|c|}
\hline
       &      & \multicolumn{2}{c|}
{Distance function } &         \\
Atomic & Total & \multicolumn{2}{c|}
{deviation from the unit} &Absolute \\
\cline{3-4}
species & energy  &The set                                   &The set                                 &error      \\
        & (a. u.) & ${\cal E}_{2m,2}$                        & $A^{-1}(2m,p,2){\cal E}_{2m,2}$        &in energy  \\
        &         &\phantom{$A^{-1}(2m,p,2){\cal E}_{2m,2}$} &                                         &(a. u.)    \\
\hline

 Li$( ^2S)$ & -7.433465  &$   <  10^{-15}$ &$ <2\cdot 10^{-8}$&$ <10^{-15}$\\
 Be$( ^1S)$ & -14.618569 &$4\cdot 10^{-13}$&$7\cdot 10^{-8}$  &$4\cdot 10^{-12}$\\
 B$(^2A)$   & -24.591900 &$4\cdot 10^{-13}$&$4\cdot 10^{-8}$  &$7\cdot 10^{-12}$\\
 C$(^3P)$   & -37.761693 &$<  10^{-15}$    &$< 10^{-15}$      &$< 10^{-15}$\\
 C$(^1D)$   & -37.707142 &$ <10^{-15}$     &$6\cdot 10^{-15}$ &$< 10^{-15}$\\
 N$(^4S)$   & -54.480030 &$< 10^{-15}$     &$1\cdot 10^{-15}$ &$< 10^{-15}$\\
 O$(^3P)$   & -74.844711 &$< 10^{-15}$     &$< 10^{-15}$      &$<10^{-15}$\\
 O$(^1D)$   & -74.765527 &$< 10^{-15}$     &$3\cdot 10^{-15}$ &$<10^{-15}$\\
 \hline
 \end{tabular}
 \end{table}

\newpage
\begin{table}[ht]
\caption{Results of simple calculations testing properties of the convex set ${\cal V}_{2m,p,2}$:
small molecules and ions}
\vspace{10mm}
\begin{tabular}{|l|c|c|c|c|c|}
\hline
       &      & \multicolumn{2}{c|}
{Distance function } &         \\
Molecular & Total & \multicolumn{2}{c|}
{deviation from the unit} &Absolute \\
\cline{3-4}
species & energy  &The set                                   &The set                                 &error      \\
        & (a. u.) & ${\cal E}_{2m,2}$                        & $A^{-1}(2m,p,2){\cal E}_{2m,2}$        &in energy  \\
        &         &\phantom{$A^{-1}(2m,p,2){\cal E}_{2m,2}$} &                                         &(a. u.)    \\
\hline
 LiH$(^1\Sigma ^+)$ & -8.016132 &$6\cdot 10^{-15}$&$7\cdot 10^{-6}$&4$\cdot 10^{-14}$\\
 CH$_2( ^3B_2)$     & -38.979862 &$4\cdot 10^{-8}$&$3\cdot 10^{-15}$&$ 6\cdot 10^{-14}$\\
 CH$_2( ^1A_1)$     & -38.922201 &$2\cdot 10^{-7}$&$5\cdot 10^{-15}$&$1\cdot 10^{-13}$\\
 NH$_2( ^2B_2)$     & -55.478526 &$6\cdot 10^{-9}$&$< 10^{-15}$&$2\cdot 10^{-14}$\\
 NH$_3(^1A_1)$      & -56.250080 &$4\cdot 10^{-8}$&$1\cdot 10^{-14}$&$3\cdot 10^{-13}$\\
 H$_2$O$( ^1A_1)$   & -76.077354 &$4\cdot 10^{-8}$&$8 \cdot 10^{-15}$&$3\cdot 10^{-13}$\\
 H$_2$O$( ^3B_2)$   & -75.796065 &$1\cdot 10^{-8}$&$1\cdot 10^{-14}$&$4\cdot 10^{-13}$\\
 H$_2$O$(^1B_2)$    & -75.765581 &$3\cdot 10^{-7}$&$1\cdot 10^{-14}$&$4\cdot 10^{-13}$\\
 H$_2$O$^+(^2A_1)$  & -75.634415 &$6\cdot 10^{-10}$&$2\cdot 10^{-15}$&$1\cdot 10^{-13}$\\
 H$_2$O$^+(^2B_2)$  & -75.560706 &$5\cdot 10^{-11}$&$2\cdot 10^{-15}$&$1\cdot 10^{-13}$\\
 \hline
 \end{tabular}
 \end{table}

\end{document}